\documentclass[aps,prd,twocolumn,nofootinbib,superscriptaddress]{revtex4-2}

\usepackage[utf8]{inputenc}
\usepackage{color}
\usepackage{graphicx}
\usepackage{amsmath}
\usepackage{amssymb}
\usepackage{bm}
\usepackage{acronym}
\usepackage{ifthen}
\usepackage{blindtext}
\usepackage[normalem]{ulem}
\usepackage{hyperref}
\hypersetup{
colorlinks=true,
linkcolor=blue,
filecolor=magenta,
urlcolor=blue,
citecolor=blue
}
\usepackage{orcidlink}
\usepackage{etoolbox}
\usepackage{fancyhdr}
\usepackage{xspace}
\usepackage{textcomp}
\usepackage{multirow}
\usepackage[caption=false]{subfig}
\usepackage{lineno}
\usepackage{orcidlink}
\usepackage{soul}
\usepackage{listings}
\usepackage[ruled,vlined]{algorithm2e}

\SetCommentSty{mycommfont}

\newtoggle{fullauthorlist}
\toggletrue{fullauthorlist}
\newtoggle{endauthorlist}
\toggletrue{endauthorlist}

\newcommand{\td}{{\rm d}}
\newcommand{\Prob}{\mathcal{P}}

\begin{document}


\title{
Performance of an  instrumented baffle placed at the entrance of Virgo’s  end mirror vacuum tower during O5
}

\author{M. Andr\'es-Carcasona\orcidlink{0000-0002-8738-1672}}
\email{mandres@ifae.es (corresponding author)}
\affiliation{Institut de Física d'Altes Energies (IFAE), The Barcelona Institute of Science and Technology, Campus UAB, E-08193 Bellaterra (Barcelona), Spain}
\author{M. Mart\'inez\orcidlink{0000-0002-3135-945X}}
\affiliation{Institut de Física d'Altes Energies (IFAE), The Barcelona Institute of Science and Technology, Campus UAB, E-08193 Bellaterra (Barcelona), Spain}
\affiliation{Catalan Institution for Research and Advanced Studies (ICREA), E-08010 Barcelona, Spain}
\author{Ll. M. Mir\orcidlink{0000-0002-4276-715X}}
\affiliation{Institut de Física d'Altes Energies (IFAE), The Barcelona Institute of Science and Technology, Campus UAB, E-08193 Bellaterra (Barcelona), Spain}
\author{J. Mundet\orcidlink{0000-0003-1091-1695}}
\affiliation{Institut de Física d'Altes Energies (IFAE), The Barcelona Institute of Science and Technology, Campus UAB, E-08193 Bellaterra (Barcelona), Spain}
\author{H. Yamamoto\orcidlink{0000-0001-6919-9570}}
\affiliation{LIGO laboratory, California Institute of Technology (Caltech), Pasadena, CA, US}

\date{\today}


\begin{abstract}

 In this article, we present results on the simulated performance of an instrumented baffle installed at the entrance of the vacuum towers hosting the end mirrors of Virgo's main Fabry-Pérot cavities.  The installation of instrumented baffles is part of the Advanced Virgo Plus upgrade in time for the O5 observing run. They were originally envisaged to be suspended, mounted on new payloads and surrounding new larger end mirrors. The current Virgo upgrade plan includes the replacement of the mirrors with new ones of better quality and same dimensions, leaving the installation of new payloads and larger end mirrors to a post-O5 upgrade phase still to be defined. 
Here we  demonstrate that placing the instrumented baffles just beyond the cryotrap gate valve and in front of the end mirrors would be equally effective for monitoring scattered light inside the cavities.  This new location, more than a meter away from the mirror, further reduces the risk of contamination and any potential interference with the mirrors, preserves the full capability to monitor scattered light, and decouples the instrumented baffle timeline from the plans for installing large mirrors in the experiment. We provide an estimate of the light distribution the baffles would encounter under both nominal and non-nominal conditions, as well as an assessment of the scattered light noise introduced by these baffles in this new location, confirming that they would not compromise Virgo's sensitivity.

\end{abstract}

\pacs{}  

\maketitle

\section{Introduction}
\label{sec:intro}
Advanced Virgo Plus (AdV+) represents the final configuration of the interferometer for the $5^{\mathrm{th}}$ observing run (O5).  After the inclusion of a frequency dependent squeezing system \cite{VIRGO:2023wes} during the Phase I of the upgrade, the main improvements expected for Phase II were based on decreasing the mechanical friction of the coating, increasing the laser power and enlarging the end mirrors (EM) to make the beam spot larger and dilute more the energy \cite{VIRGO:2023elp}. 

This proposed upgrade included the installation of new suspended payloads to support the larger {\color{black}{$\mathcal{O}$(100~kg)}} mirrors along with new instrumented baffles surrounding the optics. These baffles followed the same {{\color{black}conceptual}} design succesfully implemented during the Phase I around the input mode cleaner (IMC) end 
mirror~\cite{Romero-Rodriguez:2020bys,Ballester:2021bua,Andres-Carcasona:2022imx}. 
A study was conducted to evaluate the effectiveness of these instrumented baffles, demonstrating their ability to effectively monitor the scattered light field under both nominal and non-nominal conditions~\cite{Macquet:2022simsVirgo}. 
The results indicated that the baffle could be used to detect point absorbers on the mirrors and monitor misalignments in the cavity. 
{\color{black}{Stray light remains a critical source of noise, potentially impacting the sensitivity of the interferometer~\cite{Accadia:2010zzb,Fiori:2020arj,Was:2020ziy,LIGO:2020zwl,Longo:2020onu,Longo:2021avq,Longo:2023vac}, and an active light monitoring inside the main cavity can contribute to a better understanding of its origin and its mitigation. }}


The experienced difficulties in Virgo during the O4 commissioning work, related to operating marginally-stable recycling cavities in the experiment, together with the need for an extended R\&D period in the existing efforts for defining new optical coatings for the main mirrors, have translated into a revisited Virgo upgrade plan in which the installation of large mirrors is deferred to a post-O5 upgrade framework. The current pre-O5 upgrade plan is now focused, among other actions, on the possible installation of stable recycling cavities and the replacement of the existing main mirrors in the arms with new ones of identical dimension and improved quality, with a reduced number of defects \cite{AdV+PhaseII}. The latter is  crucial as point defects in the existing mirrors translate into significant losses and will prevent the experiment from increasing the injected laser power, thus limiting the sensitivity.   


In this paper, we explore the feasibility of placing the instrumented baffles at a different location than originally planned and evaluate whether they would meet their primary goals without limiting the sensitivity by introducing additional noise sources.

The paper is organized as follows. In Sec.~\ref{sec:baffLoc} the new proposed location of the baffle is presented and justified. In Sec.~\ref{sec:methods} the methods to assess the performance of the instrumented baffle and to compute its induced noise are explained. In Sec.~\ref{sec:baff_per} the results showing the light distribution that the baffle would measure under ideal and non-ideal conditions are presented. Finally, in Sec.~\ref{sec:scattered_light} an estimate of the scattered light noise that this baffle would introduce is presented and compared with the expected Virgo sensitivity.

This is particularly important, as the original design considered suspended 
baffles~\cite{Macquet:2022simsVirgo}, significantly reducing the level of vibrations and minimizing the scattered light noise. In contrast, the new ground-based location is subject to higher levels of vibration, which requires a new detailed study.

\section{New baffle location} 
\label{sec:baffLoc}

\begin{figure*}[htbp]
    \centering
    \includegraphics[width = \textwidth]{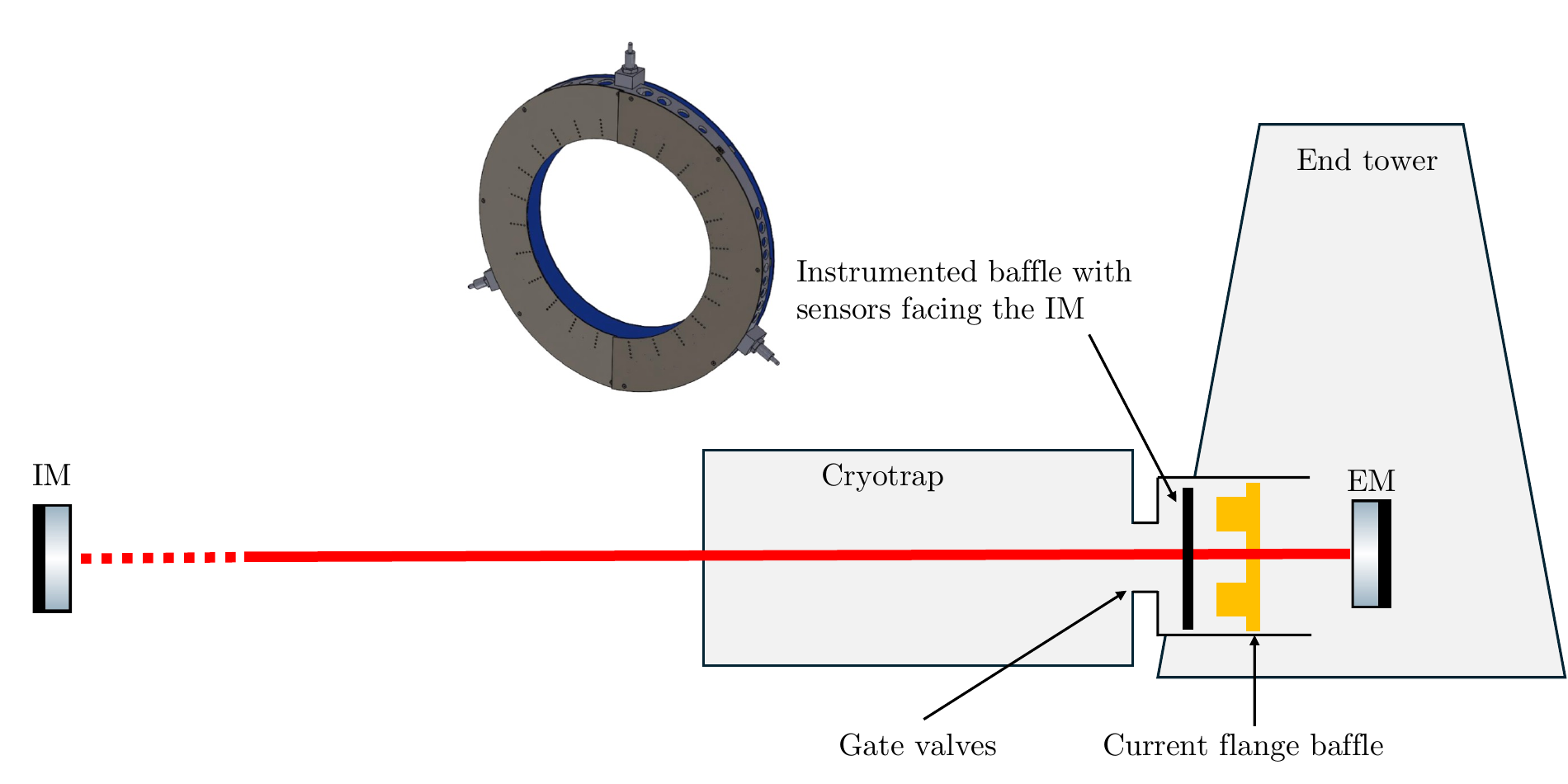}
    \caption{Sketch (not to scale) of the EM tower and cryotrap with a 3D render of the instrumented baffle and the already installed flange baffle.}
    \label{fig:Blueprint_Cryotrap}
\end{figure*}

Figure~\ref{fig:Blueprint_Cryotrap} presents a sketch of the cryotrap and vacuum tower areas surrounding the end mirror in the Fabry-Pérot cavity.  Detailed mechanical integration studies showed that the instrumented baffle could be installed in situ at the entrance of the vacuum tower, between the gate valves and the existing flange baffle, without interfering with the path of the auxiliary laser used to monitor the mirror surface\footnote{Very early studies back in 2019 totally excluded the possibility of placing the instrumented baffle inside the cryotrap area as it required a major hardware intervention.}.  In this new location, the instrumented baffle will be 1.4~m away from the mirror in the $z$-axis, along the laser beam direction, with the sensors facing the input mirror approximately 3~km away. As described below, this new location preserves the original baffle functionalities in terms of exposure to the light in the cavity.  

The design of the baffle itself remains mostly unchanged compared to that presented in Ref.~\cite{Macquet:2022simsVirgo}. The inner (outer) diameter is fixed to 0.52~m (0.8~m) and the baffle contains five concentric rings of 24 photodiodes, each located 0.27, 0.28, 0.29, 0.30 and 0.31~m away from the center. 
Each photodiode is similar to those already placed in the IMC baffle~\cite{Romero-Rodriguez:2020bys,Ballester:2021bua,Andres-Carcasona:2022imx,Macquet:2022simsVirgo} and are based on the  S13955-01 model, having a dynamic range of 13~mW and an approximate resolution of 60~$\mu$W. 
These sensors are designed to have a readout speed of 1~kHz, 
allowing for future correlations with other sensors and control channels at Virgo.

The new design includes some modifications necessary to place the baffle in the new location. As shown in Fig.~\ref{fig:Blueprint_Cryotrap}, the baffle will be fixed to the inner walls of the vacuum tube using three plungers, spaced $120^\circ$ among them.  The instrumented baffle includes encapsulated electronics and is now equipped with an additional optically coated baffle in the backside to absorb the large-angle scattered light originated in the end mirror.

\section{Scattered Light Noise Computing Methods}
\label{sec:methods}

{\color{black}{
This section presents the different tools used to perform the analysis of the scattered light noise originated by the instrumented baffle: the modeling techniques, the mirror maps used, the vibration of the baffle, and the computation of the scattered light noise. 
In this work we assume the main source of scattering is due to mirror imperfections and we consider two separate noise sources: back-scattering noise from the baffles at both small and large angles,  and diffraction noise related to the limited baffle aperture. 
}}

\subsection{Modeling tools}

{\color{black}{
We use the program \textit{Stationary Interferometer Simulations} 
(SIS)~\cite{Romero21} to compute the scattered light field  inside the cavity, the light illuminating the baffle,  and the induced noise. It is an FFT-based code that uses the paraxial approximation to compute the field inside the Fabry-P\'erot cavity given the appropriate set of parameters that define it. SIS is suitable for simulating low-angle scattering originated from the input mirror about 3~km away.  In order to compute the large-angle scattering illuminating the baffle's backside, as originated from the nearby end mirror, SIS cannot be used. The simulation of large angle scattering would require a very fine resolution of the mirror map, making it computationally unfeasible (see Sec.~\ref{sec:mirror_quality} for further details). }}

To estimate the baffle back-scattering induced noise at large angles,  an analytical approach must be used, as done in Refs.~\cite{Romero-Rodriguez:2022mje,Andres-Carcasona:2023qom}. This can be achieved by means of the bidirectional reflectance distribution function (BRDF), defined as the quotient between the differential surface radiance and the surface irradiance, which takes the form~\cite{stover2012optical}

\begin{equation}
    \mathrm{BRDF} = \frac{1}{P_i\cos \theta_s}\frac{\td P_s}{\td \Omega_s}\, ,
\end{equation}

\noindent being the subscripts $i$ and $s$ the incident and scattered quantities, respectively. 
The quantity $\td P_s/\td \Omega_s$ indicates the differential amount of power that goes into a solid angle $\td \Omega_s$, and $P_i$ is the incident power.
To estimate the BRDF, the relation with the 2D power spectral distribution (PSD) of the surface, $S(f_x,f_y)$, can be used. This is~\cite{stover2012optical}

\begin{equation} \label{eq:BRDFdef}
    \mathrm{BRDF} = \frac{16\pi^2}{\lambda^4}\cos (\theta_i) \cos (\theta_s) Q S(f_x,f_y)\, ,
\end{equation}

\noindent where $\lambda$ is the wavelength of the laser and $Q$ represents the geometrical mean of the specular reflectances measured at the incident and scattered angles. 
Since Virgo mirrors have almost perfect reflectance, we assume $Q\approx 1$. 
Additionally, we assume that the incident angle is perpendicular to the mirror, which implies that $\cos(\theta_i)\approx 1$. 
Usually the available power spectral density (PSD) is one-dimensional~\cite{HiroPSD}  
(given the cylindrical symmetry, the only relevant dimension is the radial one) 
and it is useful to modify Eq.~\eqref{eq:BRDFdef} to~\cite{stover2012optical}

\begin{equation}\label{eq:BRDF_PSD}
    \mathrm{BRDF} = \frac{16\pi^2}{\lambda^3}\cos (\theta) \mathrm{PSD}_{\mathrm{1D}}\left( \frac{\sin(\theta)}{\lambda} \right)\ , 
\end{equation}

\noindent
where we dropped the indices as all the angles now refer to the scattered ones. Therefore, 
with an available measured or estimated PSD of the mirror surface, 
we can obtain an estimation of the BRDF.

\subsubsection{Back-scattering noise}

To estimate the noise of backscattered light ($h_{\mathrm{bs}}$), the relation from Refs.~\cite{Thorne89,FlanaganThorne95_Diff,Vinet96,Vinet97,Brisson98,Andres-Carcasona:2023qom} can be used, which equals

\begin{equation}
    \tilde{h}^2_{\mathrm{bs}}(f)=\frac{1}{L^2}\left[\lambda^2+\left(\frac{8 \Gamma P_{circ}}{cM\pi f^2}\right)^{2}\right]\frac{\td \Prob}{\td \Omega_{bs}} X^2(f) K \, ,
\end{equation}

\noindent where $\Gamma$ is the gain of the cavity formed by the input mirror (IM) and signal recycling mirror (SRM), $P_{circ}$ the circulating power inside the cavity, $\td \Prob / \td \Omega_{bs}$ the BRDF of the baffle, $L$ the length of the cavity, $M$ the mass of the EM and IM, $c$ the speed of light, $X(f)$ the scatterer surface up-converted displacement (details are presented in~Sec.~\ref{sec:baff_displacement}), and

\begin{equation}\label{eq:K}
 K = 
    \frac{1}{z^2}\left( \frac{\td \Prob}{\td \Omega_{ms}}\right)^2\delta \Omega_{ms}\, ,
\end{equation}

\noindent being $z$ the distance between the scattering mirror and the position of the baffle, 
$\delta \Omega_{ms}$ the solid angle seen by the photon being scattered off the mirror and $\td \Prob/\td \Omega_{ms}$ the probability of a photon of being scattered by the mirror. This last probability is the one that will either be estimated analytically for the end mirror towards the baffle with the BRDF or simulating the cavity using SIS for the contribution from the input mirror.

\subsubsection{Diffraction noise}
The diffraction noise ($h_{\mathrm{df}}$) can be estimated by assuming a coherent, paraxial wave propagation, accounting for the four mirrors in the main cavities of the interferometer and only one baffle as in Refs.~\cite{Thorne89,FlanaganThorne95_Diff} like

\begin{equation}\label{eq:diff(t)}
    h_{\mathrm{df}}(t) = \frac{\lambda}{\pi L}\mathbb{I}\left\{  \int_0^{2 \pi} E_{IM\to B} E_{B\to EM} R X(t) \td \varphi \right\}\, ,
\end{equation}

\noindent where $E_{IM\to B}$ represents the propagated field from the IM to the baffle, $E_{B\to EM}$ the propagated field from the baffle to the EM, $R$ the inner radius of the baffle and $X(t)$ the baffle displacement in time domain. Here, $\mathbb{I}$ represents the imaginary part.

Originally, Refs.~\cite{Thorne89,FlanaganThorne95_Diff} only considered the phase noise contribution, but as mentioned in Refs.~\cite{Hiro_ETMripple,Andres-Carcasona:2023qom}, the effect of the amplitude noise is non-negligible, especially at low frequencies. Therefore, assuming a stationary field and adding the amplitude noise contribution, the strain noise in the frequency domain can be computed by taking the Fourier transform of Eq.~\eqref{eq:diff(t)} as 

\begin{equation}
    \tilde{h}_{\mathrm{df}}(f) = \sqrt{\lambda^2+\left(\frac{8 \Gamma P_{circ}}{cM\pi f^2}\right)^{2} } \frac{X(f)}{\pi L}C\, ,
\end{equation}
where 
\begin{equation}\label{eq:C}
    C = \mathbb{I}\left\{  \int_0^{2 \pi} E_{IM\to B} E_{B\to EM} R \td \varphi \right\}\, .
\end{equation}

\subsection{Mirror quality} 
\label{sec:mirror_quality}

The quality of the mirrors is one of the main variables that influences the amount of scattered light that will reach the baffle. The better the quality of the mirrors, the less scattered light is expected. This quality is dominated mainly by the coating and polishing of the surface.

\begin{figure*}[htbp]
    \centering
    \includegraphics[width=\textwidth]{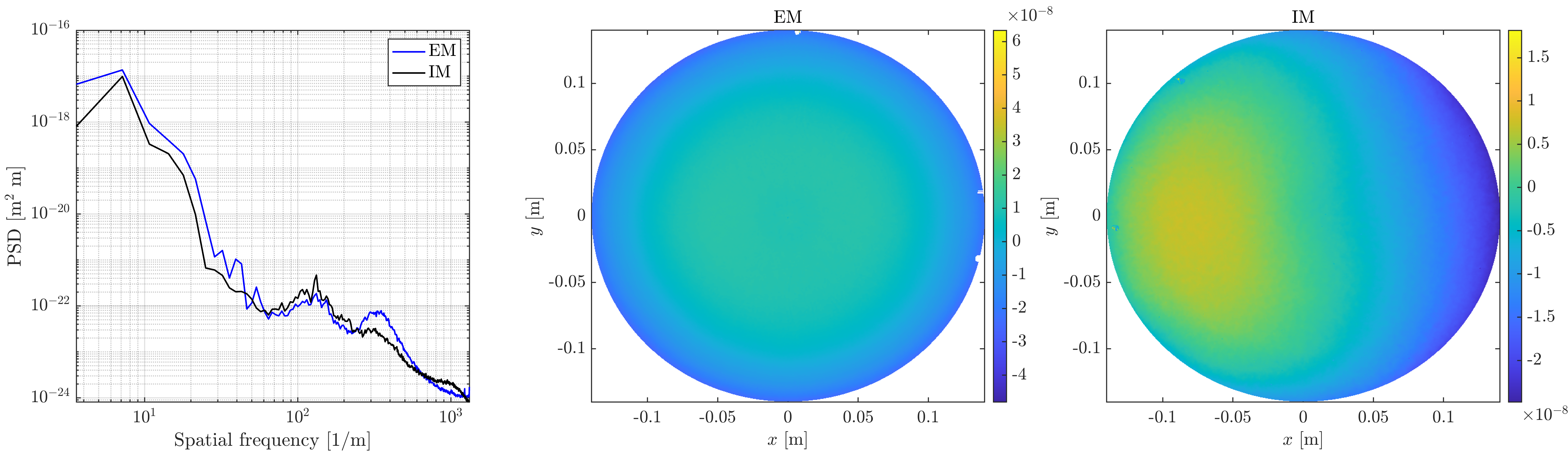}
    \caption{(\textit{left}) One-dimensional PSD of the mirror maps used. (\textit{center}) Height of the deviation from a perfect spherical surface in meters used for the end mirror. (\textit{right}) Height of the deviation from a perfect spherical surface in meters used for the input mirror. }
    \label{fig:Mmap}
\end{figure*}

The mirror map measures the deviation of the mirror surface from a perfect one and this dictates the overall scattering. For this work, we use the measured mirror maps for Virgo's O3 mirrors \cite{Degallaix:2019cta}. These are shown, alongside their one-dimensional PSD in Fig.~\ref{fig:Mmap}. These mirror maps have a limitation dictated by their resolution. The scattering angles that can be inferred are $\theta\sim \lambda \xi$, where $\xi$ is the one-dimensional spatial frequency, but only the ones in the range $\xi\in[R_m^{-1},(2\Delta)^{-1}]$ can be resolved \cite{Andres-Carcasona:2023qom}. The lower limit corresponds to a physical restriction of not being able to account for larger wavelengths than the radius of the mirror itself, $R_m$. The upper limit is related to the Nyquist frequency and is set by the resolution of the mirror map, $\Delta$. 

Therefore, the side of the baffle that is facing the  input mirror can be simulated, as it has a small scattering angle, but the back-side of the baffle cannot, as the finite resolution of the mirror map is unable to resolve such large angles. 

The value of the BRDF obtained evaluating Eq.~\eqref{eq:BRDF_PSD}  for the angle from the end mirror to the back side of the baffle, 
which is $\theta\approx 0.26/1.4 \approx 0.19$~rad, and the PSD from Fig.~\ref{fig:Mmap},
is approximately $10^{-6}$~str$^{-1}$. 
This value is consistent with the high-angle scattering measurements presented in Ref.~\cite{Romero-Rodriguez:2022mje}.

\subsection{Baffle displacement}\label{sec:baff_displacement}
The scattered light induced by the baffle is an inevitable consequence of its vibrations and this is why they appear in the equations to estimate this noise in the $X(f)$ term. We use the data of a velocimeter and an accelerometer close to the cryotrap area, as done in Ref.~\cite{Romero-Rodriguez:2022mje}. We take $1200$~s and $30$~s of data starting at a GPS time of $1400360418$~s for the velocimeter and accelerometer, respectively. The former is sampled at 1~kHz and the latter at 10~kHz. This time period is chosen because there was a high microseism activity, at the 99\% percentile level.
These data are then low-filtered for the velocimeter and high-filtered for the accelerometer with a cutoff frequency of 50~Hz and converted into the frequency domain to express both of them as displacement noise. This is done by recalling that

\begin{equation}
    \mathrm{PSD}_{\rm acc}(f) = (2\pi f)^2  \mathrm{PSD}_{\rm vel}(f) = (2\pi f)^{4}  \mathrm{PSD}_{\rm dis}(f)~.
\end{equation}

The final result is dominated by the velocimeter data at low frequencies and by the accelerometer one at high frequencies, as the measurements of each sensor are only reliable in a region of frequencies.

\begin{figure}[htbp]
    \centering
    \includegraphics[width = \columnwidth]{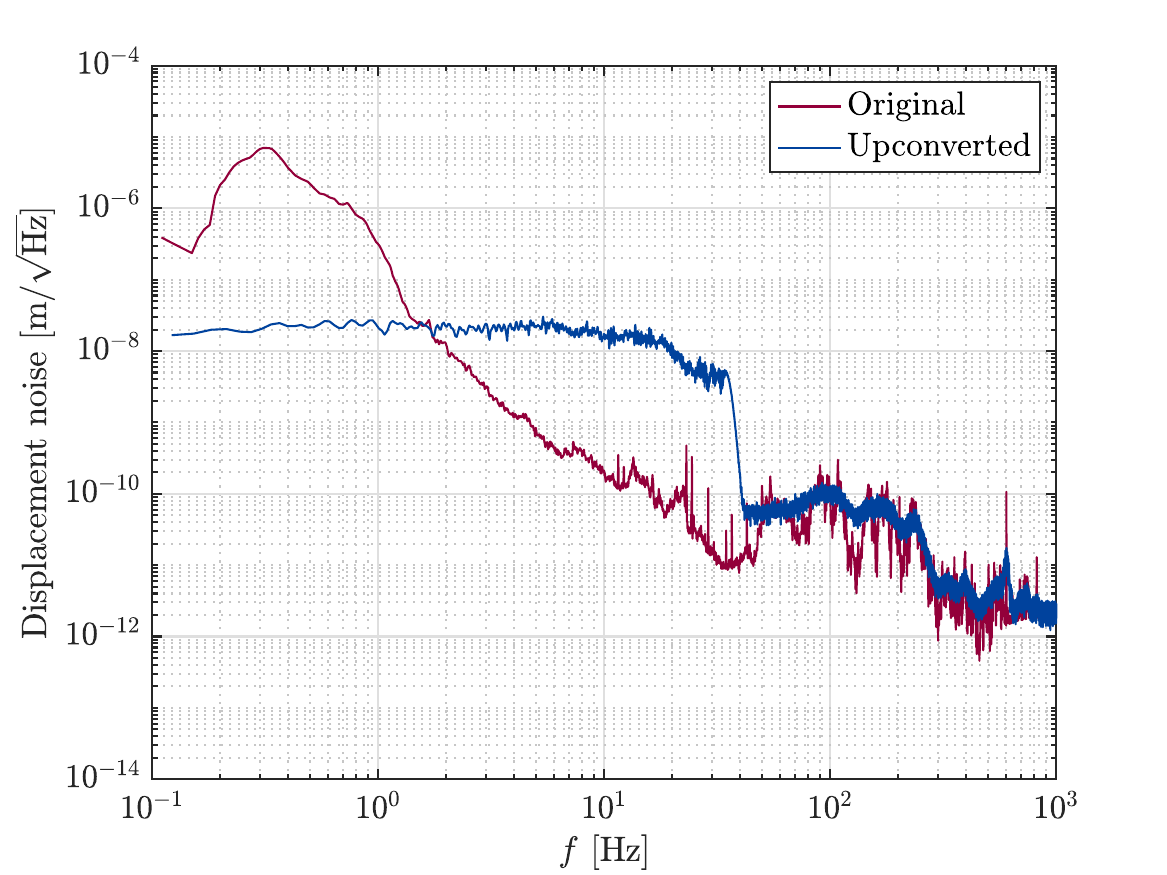}
    \caption{Spectrum of the original data and spectrum resulting from the up-conversion procedure for the displacement data.}
    \label{fig:velocimeter_PSD}
\end{figure}

These data must be up-converted to account for the phenomenon known as phase-wrapping. We follow the procedure thoroughly described in Refs.~\cite{Andres-Carcasona:2023qom,CE_backscattering} and the resulting spectrum, alongside the original raw data, are displayed in Fig.~\ref{fig:velocimeter_PSD}. 
As expected, the up-conversion has the effect of reducing the low frequency noise to increase the high-frequency one.

\section{Expected light distribution} \label{sec:baff_per}
\begin{figure*}[htbp]
    \centering
    \includegraphics[width=\textwidth]{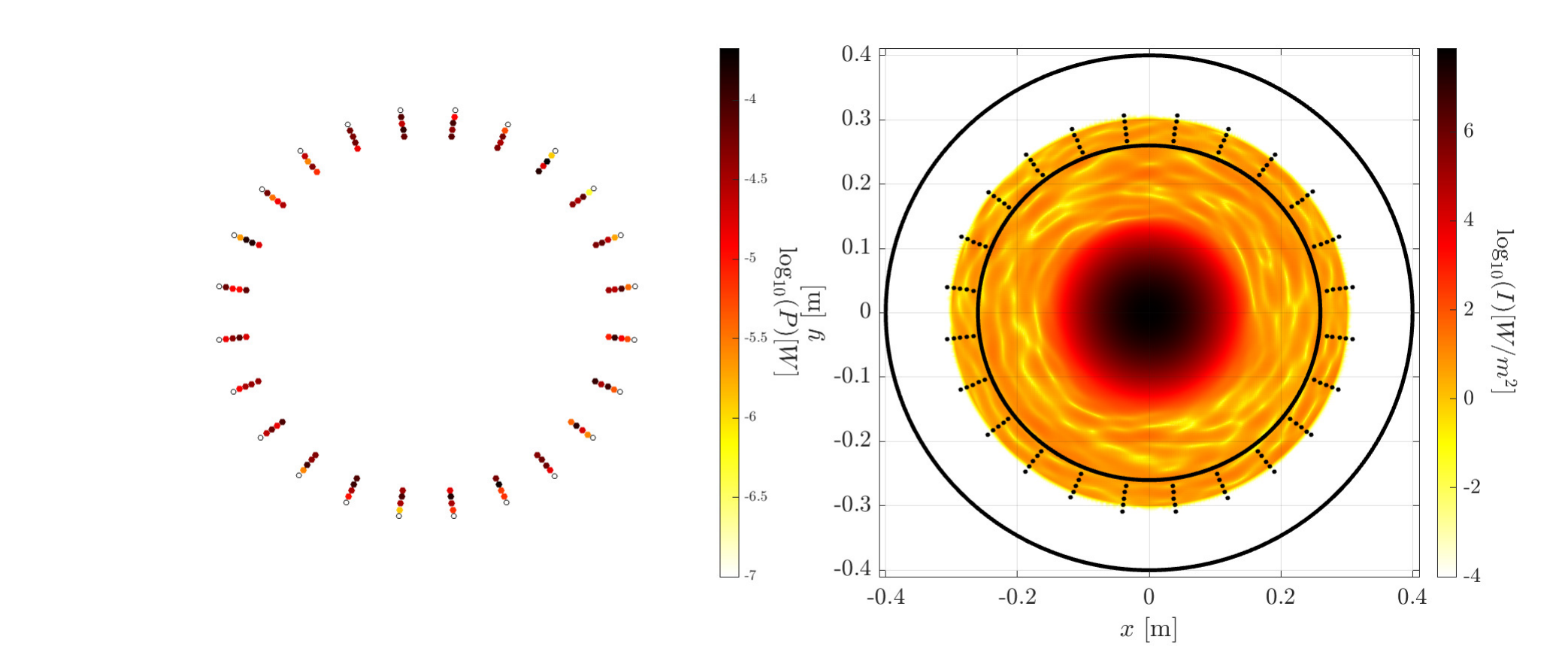}
    \caption{(\textit{left}) Power reaching each photodiode of the instrumented baffle. (\textit{right}) Field intensity that reaches the instrumented baffle. The black circles indicate the limits of the baffle.}
    \label{fig:lightdistribution}
\end{figure*}

Following closely the work in Ref.~\cite{Macquet:2022simsVirgo},  we evaluate the scenario of placing an instrumented baffle between the gate valves and the end mirror. This study is motivated by the fact that, in this new location, the stray light reaching the baffle could be affected by the limited apertures at the entrance of the vacuum tower. Moreover, the beam optics differs from that originally anticipated for O5, with a smaller beam size at the end mirror \cite{AdV+PhaseII}, potentially leading to reduced signals in the baffle sensors.  
We first estimated the distribution of light illuminating the baffle in the new position using SIS, the parameters displayed in Tab.~\ref{tab:parameters}, and the mirror maps described in Sec.~\ref{sec:mirror_quality}. The resulting simulated field distribution is presented in Fig.~\ref{fig:lightdistribution}.

\begin{table}[htb]
\begin{center}
\footnotesize
\begin{tabular}{c |c | c}
\hline \hline
         Variable & Description & Value \\ \hline \hline 
        $\lambda$ & Laser wavelength & $1064$ nm \\ \hline
        $P$ & Input power & $1.364$ kW \\ \hline
        $L$ & Cavity length & $3$ km \\ \hline
        $R_m$ & Radii of the mirrors & $16.5$~cm \\ \hline
        $\mathcal{R}_{\mathrm{IM}}$ & Radius of curvature IM & $1420$ m \\ \hline
        $\mathcal{R}_{\mathrm{EM}}$ & Radius of curvature EM & $1683$ m \\ \hline
        $w_{\mathrm{IM}}$ & Beam spot at IM & $4.87$ cm \\ \hline
        $w_{\mathrm{EM}}$ & Beam spot at EM & $5.80$ cm \\ \hline
        $T_{\mathrm{IM}}$ & Transmissivity IM & $0.01375$  \\ \hline
        $T_{\mathrm{EM}}$ & Transmissivity EM & $5\times 10^{-6}$  \\ \hline
        $t_{\mathrm{IM}}$ & Thickness IM & $20$~cm  \\ \hline
        $t_{\mathrm{EM}}$ & Thickness EM & $20$~cm  \\ \hline \hline
    \end{tabular}
    \caption{Parameters of the Fabry-P\'erot cavity used in this work. They corresponds to the ones in Ref.~\cite{VIRGO:2014yos}, except the input power which is chosen to match the circulating power of O5~\cite{VIRGO:2023elp}.}
    \label{tab:parameters}
    \end{center}
\end{table}

Considering nominal conditions, the simulation indicates that the baffle will be sensitive to the scattered light field inside the cavity. As expected, since the distance between the original position near the mirrors and the new position at the entrance of the vacuum tower is less than 2~m, the results are very similar to those presented in Ref.~\cite{Macquet:2022simsVirgo}. Therefore, the new position does not diminish the capabilities of monitoring the scattered light field distribution. 

The first four inner rings, closer to the beam axis, will be exposed to light power in the range between $10^{-5}$ and $10^{-4}$~W, well within the dynamic range and resolution of the sensors. In the case of the most outer ring, it is not expected to receive light in nominal conditions, since it is geometrically shielded by the cryobaffle and the gate valve limited apertures. However, it is designed for the purpose of measuring the light distribution in the cavity in the presence of large beam mis-alignments,  in a scenario in which the response of the sensors in the  inner layers will saturate. 

As shown  in Fig.~\ref{fig:lightdistribution}, the power distribution is not isotropic, with the aberrations on the mirrors surfaces generating random patterns. In principle, they can be measured by the instrumented baffle, as it was already the case in the IMC~\cite{Romero-Rodriguez:2020bys,Ballester:2021bua}.

We turn now into considering special conditions assuming misalignments in the beam in the form of either a beam offset and/or a tilt. Such offsets and tilts might be needed to avoid hitting mirror imperfections with the beam, thus decreasing the circulating power, changing the radius of curvature and increasing the scattered light noise. Altogether, this will result into changes in the light measured by the instrumented baffle.

\begin{figure*}[htbp]
    \centering
    \includegraphics[width=\textwidth]{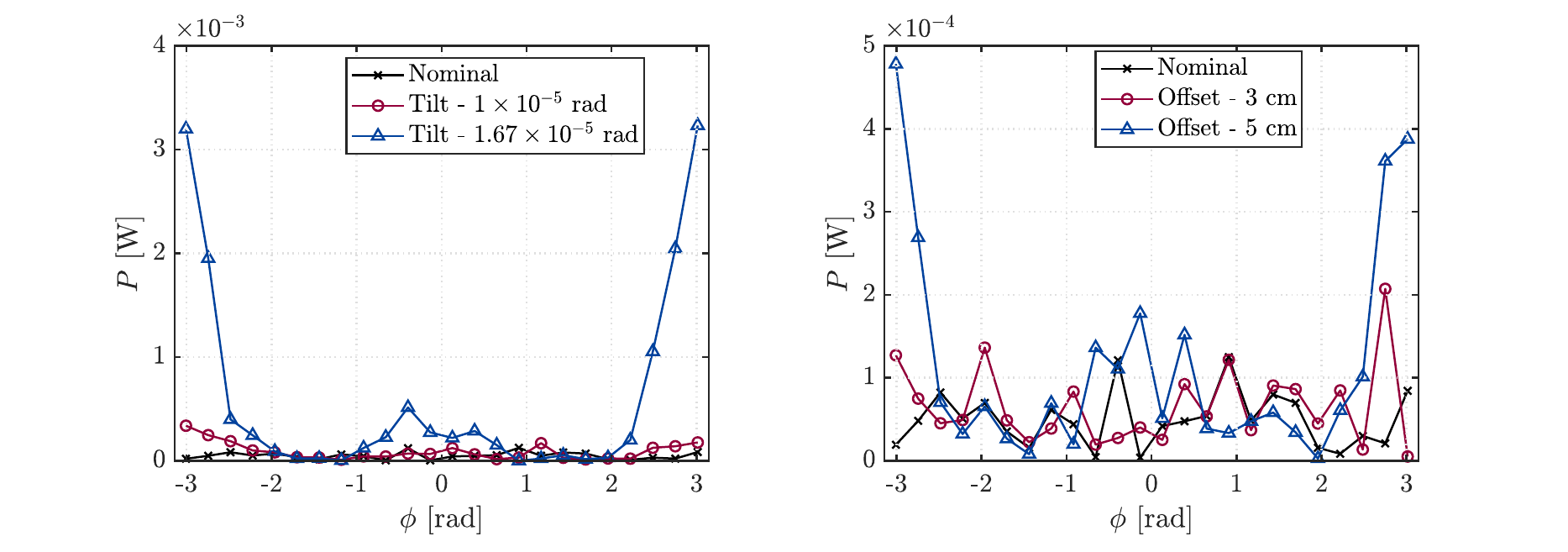}
    \caption{(\textit{left}) Power reaching the photodiodes of the inner ring for different values of the tilt angle of the beam in the $x$ direction. (\textit{right}) Power reaching the photodiodes of the inner ring for different values of the offset value of the beam in the $x$ direction.}
    \label{fig:nonideal}
\end{figure*}

Figure~\ref{fig:nonideal} shows the light power distribution measured by the sensors in the most inner ring of the instrumented baffle in nominal conditions and in an scenario with misalignments of $1\times 10^{-5}$~rad and $1.67\times 10^{-5}$~rad in the negative $x$ axis, or in the presence of beam off-sets of 3~cm and 5~cm towards the negative $x$ axis.  The magnitude of the misalignment selected falls in the range of reasonable values expected  for the interferometer. The change in the response of the sensors demonstrates that, as it was already reported in Ref.~\cite{Macquet:2022simsVirgo}, the baffle retains the capacity of detecting non-ideal conditions of the cavity and contribute to the pre-alignment of the cavity and to guide the laser beam towards the center of the mirrors.


\section{Scattered light noise} \label{sec:scattered_light}

{\color{black}{
Following the methods described in Sec.~\ref{sec:methods} and the parameters listed in Tables~\ref{tab:parameters} and~\ref{tab:parameters_2}, we have computed the induced stray light noise by the instrumented baffle in its new location. We consider both the scattered light noise from the back-scattering and from the diffraction of the baffle. }}

\begin{table}[htbp]
    \begin{center}
    \footnotesize
    \begin{tabular}{c|c|c}
    \hline \hline
         Variable & Description & Value \\ \hline \hline 
        $\Gamma$ & IM-SRM cavity gain & 15.7 \\ \hline
        $\td \Prob / \td \Omega_{bs}$ & BRDF of the baffle & $10^{-4}$ str$^{-1}$ \\ \hline
        $M$ & Mass of the mirrors & $40$ kg \\ \hline
        $P_{circ}$ & Circulating power & $392$ kW \\ \hline
        $R$ & Inner radius of the baffle & $26$~cm \\ \hline \hline
    \end{tabular}
    \caption{Additional parameters used for the estimation of the scattered light noise.}
    \label{tab:parameters_2}
    \end{center}
\end{table}

{\color{black}{
From the SIS simulation, we obtain the values $K= 1.55\times10^{-11}$ m$^{-2}$ and $C= 5.1\times 10^{-12}$ for the front side of the baffle, facing the input mirror in the cavity at a distance of about 3~km. In the case of  the back-propagating field, which would encounter the back side of the baffle, we can estimate the value of $K$ analytically using Eq.~\eqref{eq:K} with $z = 1.4$ m, $\td \Prob / \td \Omega_{ms}\approx 10^{-6}$~str$^{-1}$ and $\delta\Omega_{ms}\approx 2\pi\theta\Delta\theta\approx 0.115$~str. This yields $K = 6.0\times 10^{-14}$ m$^{-2}$. This small value, three orders of magnitude smaller than the forward scattering, implies that the noise will be dominated by the small angle scattering. }}

\begin{figure*}[htbp]
    \centering
    \includegraphics[width=\textwidth]{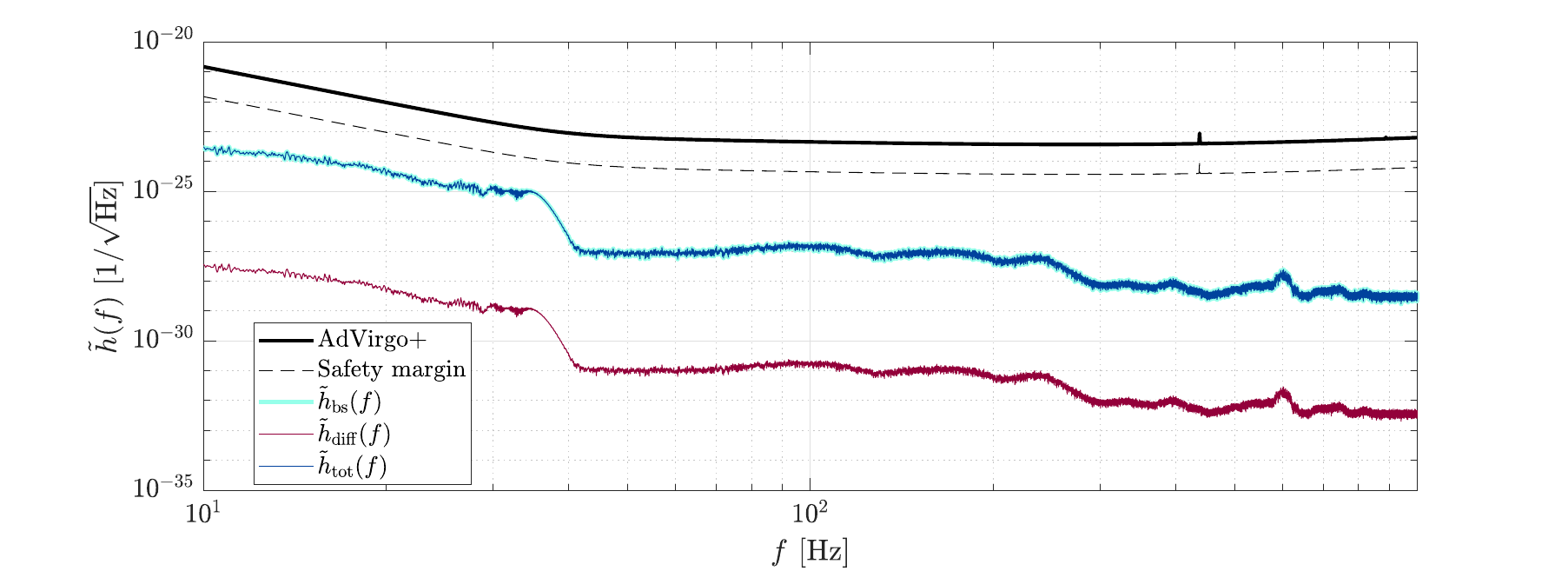}
    \caption{Projected noise spectrum of the back-scattering of the baffle, the diffraction and the total. The black solid line indicates the sensitivity of Virgo in O5 and the dashed one is one order of magnitude below this one.}
    \label{fig:noise}
\end{figure*}

{\color{black}{
Figure~\ref{fig:noise} shows the computed stray light induced noise compared to the anticipated O5 Virgo sensitivity~\cite{KAGRA:2013rdx}, where a safety margin of one order of magnitude below the projected noise level has been established, as in Refs.~\cite{Romero-Rodriguez:2022mje,Andres-Carcasona:2023qom}.}} 

\noindent

In this case, the scattered light noise induced by the baffle is dominated by the back-scattering and the effect of the diffraction from the instrumented baffle is negligible~\footnote{As discussed in Ref.~\cite{Andres-Carcasona:2023qom}, diffraction noise becomes relevant when there are multiple baffles involved and the effect is coherent.}.  As presented in Fig.~\ref{fig:noise},  the projected total scattered light noise is below the safety margin, indicating that placing the baffle in the new location at the entrance of the vacuum tower would not compromise Virgo's sensitivity.

Finally, the results in Fig.~\ref{fig:noise} do not include possible effects from the mechanical couplings of the baffle.  A careful study was carried out since the presence of resonant frequencies and a larger than one transfer factor between the motion of the cryotrap and the one of the baffle could potentially enhance the baffle vibrations and modify the spectrum used in the noise calculations. A detailed discussion on resonances and amplification factors is collected in Appendix~\ref{app:2} for the current mechanical design. We find that the inclusion of the mechanical couplings does not modify the conclusions since a single narrow resonance at 112~Hz is observed with an amplification factor,  as computed from ground to the baffle, of about 60, which does not compromise the sensitivity of the interferometer. 

\section{Conclusions }

We presented the simulated performance of an instrumented baffle installed at the entrance of the vacuum towers hosting the end mirrors of Virgo’s main Fabry-P\'erot cavities. The results indicate the baffle would be able to monitor the  scattered light inside the cavities and would have the sensitivity to detect large laser beam mis-alignments. We computed the scattered light noise induced by the baffle in this location, considering also potential enhancements due to the presence of mechanical resonances  in the baffle. Both the back-scattering and diffraction noises are estimated and are orders of magnitude below the expected sensitivity of Virgo during O5. This study  confirms that the inclusion of the instrumented baffle would not compromise Virgo’s sensitivity.

\section*{Acknowledgements}

This project has received funding from the European Union’s Horizon 2020 research and innovation programme under the Marie Skłodowska-Curie Grant Agreement No. 754510.
This work
is partially supported by the Spanish MCIN/AEI/10.13039/501100011033 under the Grants
No. SEV-2016-0588, No. PGC2018-101858-B, No. PID2020-113701GB-I00, and No. PID2023-146517NB-I00 some
of which include ERDF funds from the European Union, and by the MICINN with funding from the European Union NextGenerationEU (PRTR-C17.I1) and by the Generalitat de
Catalunya. IFAE is partially funded by the CERCA program of the Generalitat de Catalunya.
MAC is supported by the 2022 FI-00335 grant. Part of this material is based upon work supported by NSF’s LIGO Laboratory which is a major facility fully funded by the National Science Foundation and operates under Cooperative Agreement No. PHY-1764464.

\appendix

\section{Numerical transfer functions}\label{app:2}

As already mentioned, the scattered light noise induced by the baffle might be affected by the presence of mechanical resonances at given frequencies and larger than one mechanical transfer factors from ground to the baffle surface, potentially amplifying the baffle vibrations. 

Two modal analyses are performed using the finite element method (FEM) to determine the different vibration modes of the baffle, which depend on its detailed design and how it is attached to the tube. 
As shown in Fig.~\ref{fig:Blueprint_Cryotrap}, the baffle includes three plungers, spaced $120^\circ$ apart, which behave as point contacts.  The transfer function between the base of the tube and the attachment of the baffle to the tube is assumed to be one. For simplicity, the details on the electronic boards are not included but their weight is added to the instrumented baffles. 
Finally, a  global interaction between components is considered such that there is a bonding interface between all the surfaces in contact.

The first FEM analysis provides the main resonance frequencies, 
together with their equivalent mass, as collected in Tab.~\ref{tab:modes}.
The first natural frequency occurs at 112~Hz with a 92.3\% equivalent mass in the axial ($y$) direction.

\begin{table}[htbp]
    \centering
        \footnotesize
    \begin{tabular}{c|c|c|c|c}
         \hline \hline
         Mode &  $f$ [Hz] &  $x$  & $y$ & $z$\\
         \hline
1                & 112.0                   & 0.0\%                   & 92.3\%                  & 0.0\%                   \\ \hline
2                & 121.5                   & 0.0\%                   & 3.5\%                   & 0.8\%                   \\ \hline
3                & 129.2                   & 0.0\%                   & 0.8\%                   & 0.0\%                   \\ \hline
4                & 191.5                   & 0.0\%                   & 4.7\%                   & 0.1\%                   \\ \hline
5                & 243.5                   & 2.5\%                   & 0.0\%                   & 0.0\%                   \\ \hline
6                & 248.5                   & 0.2\%                   & 0.6\%                   & 0.3\%                   \\ \hline
7                & 257.5                   & 0.3\%                   & 0.0\%                   & 0.2\%                   \\ \hline
8                & 267.0                   & 0.2\%                   & 0.0\%                   & 0.0\%                   \\ \hline
9                & 273.6                   & 0.0\%                   & 0.0\%                   & 0.5\%                   \\ \hline
10               & 292.9                   & 0.3\%                   & 0.1\%                   & 52.0\%                  \\ 
         \hline \hline
    \end{tabular}
    \caption{Numerical results obtained for the various vibration modes of the instrumented baffle in the $x$, $y$ and $z$ directions. The percentage represents the mass participation in each direction.}
    \label{tab:modes}
\end{table}

\begin{figure}[htpb]
    \centering
    \includegraphics[width=0.49\textwidth]{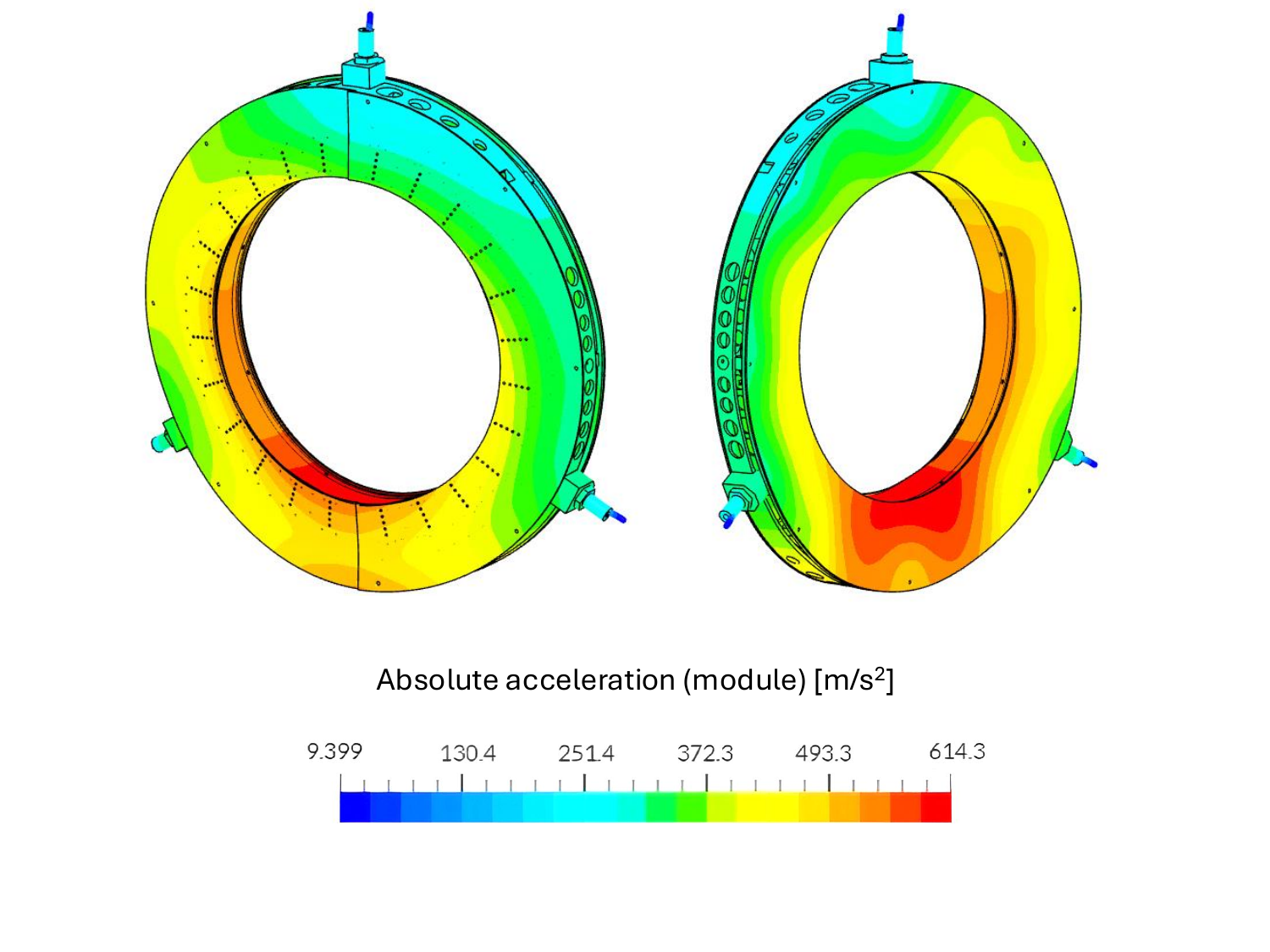}
    \caption{3-dimensional representation of the modulus of the $y$ component of the acceleration (in m/s$^2$) of the baffle when a periodical sinusoidal acceleration in the $y$ direction is applied. }
   \label{fig:3Dharmonic}
\end{figure}

We carried out an harmonic analysis around the dominant resonance at 112~Hz in the range between 1~Hz and 200~Hz, with the aim to detect other close-by resonant frequencies with a large amplification factor. To this end, 
the steady-state linear structural response to periodic sinusoidal displacements in the $x$, $y$ and $z$ directions is calculated.  Figure~\ref{fig:3Dharmonic} shows the 3-dimensional distribution of the magnitude of the $y$ component of the acceleration in the baffle structure when a periodical sinusoidal excitation in the $y$ direction of $9.8$~m/s$^2$ and $112$~Hz is applied. 

For each simulation, the ratio of the peak output acceleration over the input acceleration on each axis 
provides the amplification factor for that axis. Using this linear method, the obtained amplification factor can be used to calculate any other output displacement or acceleration for any given input,  and calculate the oscillation of the instrumented baffle for a given real seismic oscillation input. Table~\ref{tab:amplification} presents the results of the harmonic analysis for the points of the baffle that receive the maximum amplification.

\begin{table}[htbp]
    \centering
        \footnotesize
    \begin{tabular}{c c|c|c|c}
\hline \hline
  & & \multicolumn{3}{c}{Output axis} \\ 
 & & $x$ & $y$ & $z$ \\ \hline
\multirow{3}{*}{Input axis} & $x$ & 3.9 & 1.9 & 3.3 \\ \cline{2-5} 
 & $y$ & 6.3 & 62.6 & 14.9 \\ \cline{2-5} 
 & $z$ & 3.6 & 16.8 & 5.1 \\ \hline \hline
\end{tabular}
    \caption{Simulated amplification factor after the excitation of the baffle in each direction, computed as the ratio between the value obtained and the gravity acceleration of 9.8~m/s$^2$. }
    \label{tab:amplification}
\end{table}

\noindent

\begin{figure}[htbp]
    \centering
    \includegraphics[width=0.9\columnwidth]{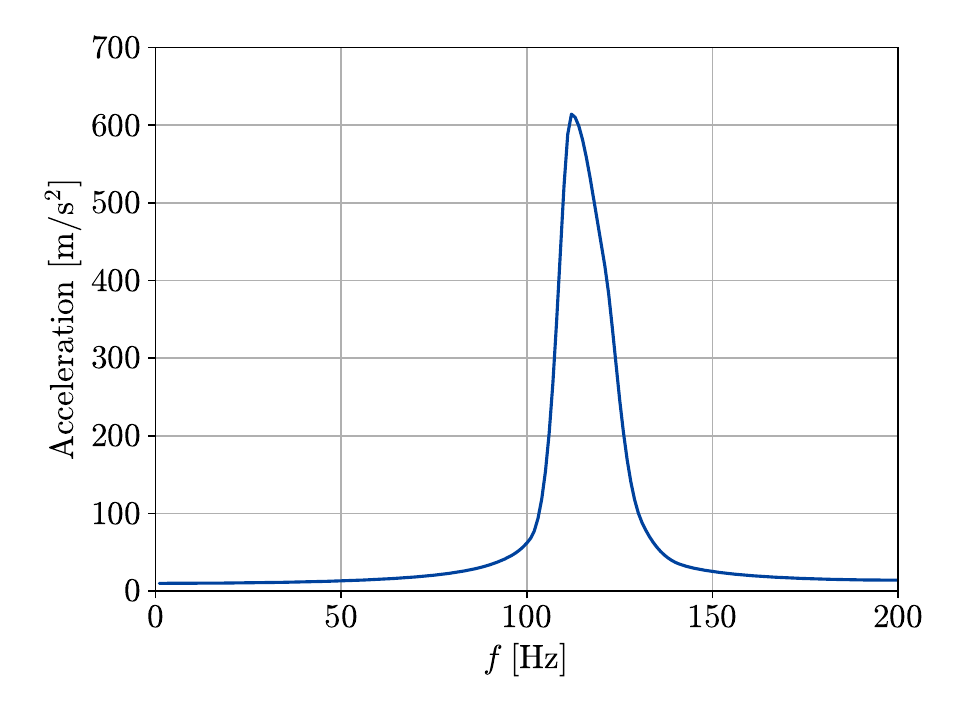}
    \caption{Frequency response for an input oscillation in the $y$ axis  and for frequencies below 200~Hz. The observed plateau corresponds to a gravity acceleration of 9.8~m/s$^2$.}
    \label{fig:transferFactor}
\end{figure}

Figure~\ref{fig:transferFactor} shows the corresponding frequency response for an input oscillation in the $y$ axis. A maximum absolute acceleration of 614 m/s$^2$ is observed in the $y$-direction for the resonant frequency of 112~Hz, corresponding to an amplification factor of 62.6.

As shown in Fig.~\ref{fig:noise}, at the frequency of 112~Hz the scattered light induced noise remains more than two orders of magnitude below the safety margin in the Virgo projected sensitivity, which indicates that the inclusion of mechanical resonances does not modify the conclusion, and the presence of the instrumented baffle will not compromise the sensitivity of the experiment.


\newpage
\bibliography{ref.bib}{}


\end{document}